\documentclass[twocolumn]{autart}    
\usepackage{graphicx}          
\usepackage{graphicx}      
\usepackage{natbib}        

\newtheorem{rmk}{Remark}
\newtheorem{exm}{Example}

\usepackage{amsmath,amssymb}
\usepackage{mathrsfs}
\usepackage[all]{xy}

\usepackage{color}

\begin{document}
\begin{frontmatter}
\vspace{-0.2cm}
\title{\vspace{-0.2cm} Explicit Simplicial Discretization of Distributed-Parameter Port-Hamiltonian Systems}
\vspace{-0.9cm}
\author[A1]{Marko Seslija}\ead{Marko.Seslija@esat.kuleuven.be}, 
\author[A]{Jacquelien M.A. Scherpen}\ead{J.M.A.Scherpen@rug.nl}, 
\author[B]{Arjan van der Schaft}\ead{A.J.van.der.Schaft@rug.nl}
\address[A1]{Department of Electrical Engineering, Katholieke Universiteit Leuven, Kasteelpark Arenberg 10,
B-3001 Leuven, Belgium}
\address[A]{Institute for Technology, Engineering and Management, University of Groningen, Nijenborgh 4, 9747 AG Groningen, The~Netherlands}
\address[B]{Johann Bernoulli Institute for Mathematics and Computer Science, University of Groningen, Nijenborgh 9, 9747 AG Groningen, The Netherlands}\vspace{-0.1cm}

\begin{abstract}
Simplicial Dirac structures as finite analogues of the canonical Stokes-Dirac structure, capturing the topological laws of the system, are defined on simplicial manifolds in terms of primal and dual cochains related by the coboundary operators. These finite-dimensional Dirac structures offer a framework for the formulation of standard input-output finite-dimensional port-Hamiltonian systems that emulate the behavior of distributed-parameter port-Hamiltonian systems. This paper elaborates on the matrix representations of simplicial Dirac structures and the resulting port-Hamiltonian systems on simplicial manifolds. {{Employing these representations, we consider the existence of structural invariants and demonstrate how they pertain to the energy shaping of port-Hamiltonian systems on simplicial manifolds.}}\vspace{-0.4cm}
\end{abstract}
\vspace{-0.4cm}
\begin{keyword}
Port-Hamiltonian systems, Dirac structures, distributed-parameter systems, structure-preserving discretization, discrete geometry\vspace{-0.2cm}
\end{keyword}\vspace{-0.4cm}

\end{frontmatter}

\section{Introduction}\vspace{-0.2cm}
A wide class of field theories can be treated as port-Hamiltonian systems \cite{vdSM02}, \cite{SchSch}. The Stokes-Dirac structure defined in \cite{vdSM02} is an infinite-dimensional Dirac structure which provides a theoretical account that permits the inclusion of varying boundary variables in the boundary problem for partial differential equations. From an interconnection and control viewpoint, such a treatment of boundary conditions is essential for the incorporation of energy exchange through the boundary, since in many applications the interconnection with the environment takes place precisely through the boundary. For numerical integration, simulation and control synthesis, it is of paramount interest to have finite-dimensional approximations that can be interconnected to one another.\vspace{-0.2cm}

Most of the numerical techniques emanating from the field of numerical analysis, however, fail to capture the intrinsic system structures and properties, such as symplecticity, conservation of momenta and energy, as well as differential gauge symmetry. Mixed finite element methods can be constructed in a such a manner that a number of important structural properties are preserved \cite{Bosavit}, \cite{Hirani}, \cite{Hiptmair}. Most of the efforts have been focused on systems on manifolds without boundary or zero energy flow through the boundary. In \cite{Golo} a mixed finite element scheme for structure-preserving discretization of port-Hamiltonian systems was proposed. The construction is clear in a one-dimensional spatial domain, but becomes complicated for higher spatial domains. Furthermore, the geometric content of the discretized variables remains moot, in sense that, for instance, the boundary variables do not genuinely live on the geometric boundary.\vspace{-0.3cm}

Recently in \cite{SeslijaArXiv}, we suggested a discrete exterior geometry approach to structure-preserving discretization of distributed-parameter port-Hamiltonian systems. The spatial domain in the continuous theory represented by a finite-dimensional smooth manifold is replaced by a homological manifold-like simplicial complex and its circumcentric dual. The smooth differential forms, in discrete setting, are mirrored by cochains on the primal and dual complexes, while the discrete exterior derivative is defined to be the coboundary operator. Discrete analogues of the Stokes-Dirac structure are the so-called simplicial Dirac structures defined on spaces of primal and dual discrete differential forms. These finite-dimensional Dirac structures offer a natural framework for the formulation of finite-dimensional port-Hamiltonian systems that emulate their infinite-dimensional counterparts. The resulting port-Hamiltonian systems are in the standard \emph{input-output} form, unlike in \cite{Golo}, where the discretized models are \emph{acausal} (given by a set of differential and algebraic equations). The explicit input-output form obtained by our scheme has the advantage from both numerical and control perspective over the implicit model presented in \cite{Golo}.\vspace{-0.2cm}

In this paper, we address the issue of matrix representations of simplicial Dirac structures by representing cochains by their coefficient vectors. In this manner, all linear operator from the continuous world can be represented by matrices, including the Hodge star, the coboundary and the trace operator. Firstly, we recall the definition of the Stokes-Dirac structure and port-Hamiltonian systems. In the third section, we define some essential concepts from discrete exterior calculus as developed in \cite{Desbrun1}, \cite{Hirani}. In order to allow the inclusion of nonzero boundary conditions on the dual cell complex, in \cite{SeslijaArXiv} we have adapted a definition of the dual boundary operator that leads to a discrete analogue of the integration-by-parts formula, which is a crucial ingredient in establishing simplicial Dirac structures on a primal simplicial complex and its circumcentric dual. We demonstrate how these simplicial Dirac structures relate to the spatially discretized wave equation on a bounded domain and to the telegraph equations on a segment. Towards the end of the paper, we consider the existence of structural invariants, which are crucial for the control by energy shaping.\vspace{-0.2cm}

{{\textbf{Goal and contributions.} This paper is written with several purposes in mind.\vspace{-0.3cm}
\begin{itemize}

\item The essential theoretical results of this paper pertaining to structure-preserving discretization, namely, Section \ref{Sec3}, \ref{Sec4} and \ref{Sec5} have been already reported in \cite{SeslijaCDC,SeslijaMathMod,SeslijaArXiv} in an algebraic topology setting. The results in this paper do not lean onto the heavy nomenclature of algebraic topology, but instead emphasizes matrix representations, making it more accessible and easier to implement. We demonstrate that a discrete differential modeling approach to consistent discretization of distributed-parameter systems is quite approachable---and, in fact, is often much simpler than its continuous counterpart.

\item We aim to render the theoretic foundation of our exposition accessible to control theorists, and the paper as such serves as a segue to the rich literature on the subject. 


\item Another contribution of this paper is given in Section~\ref{Sec7} and \ref{Sec:ECmethod}. Here we address the existence of dynamical invariants for the obtained spatially discrete systems and look at the energy-Casimir method for energy shaping. We anticipate that this line of research will lead to more elaborate and fruitful control strategies for distributed systems.

\item We hope that by the end of the paper it will become clear that the discrete geometry-based approach to modeling is \emph{not} only tied to the discretization of infinite-dimensional systems, but, instead, stands as a potent language for the system and control community.
\end{itemize}
}}

\vspace{-0.2cm}
\section{Background of port-Hamiltonian systems}\vspace{-0.2cm}
Dirac structures were originally developed in \cite{Courant}, \cite{Dorfman} as a generalization of symplectic, presymplectic and Poisson structures. Later, Dirac structures were employed as the geometric formalism underpinning generalized interconnected and constrained Hamiltonian systems \cite{L2gain}, \cite{vdSM02}.

\subsection{Dirac structures}
Let $\mathcal{X}$ be a manifold and define a pairing on $T\mathcal{X} \oplus T^\ast \mathcal{X}$ given by\vspace{-0.2cm} 
\begin{equation*}
	\left<\!\left<(f_1, e_1), (f_2, e_2) \right>\!\right> = 
	\langle e_1| f_2\rangle + \langle e_2| f_1\rangle.
\end{equation*}
For a subspace $\mathcal{D}$ of $T\mathcal{X} \oplus T^\ast \mathcal{X}$, we define the orthogonal complement $\mathcal{D}^\perp$ as the space of {{all $(f_1, e_1)$ such that $\left<\!\left<(f_1, e_1), (f_2, e_2) \right>\!\right> = 0$ for all $(f_2, e_2)$.}}  A \emph{\textbf{Dirac structure}} is then a subbundle $\mathcal{D}$ of $T\mathcal{X} \oplus T^\ast \mathcal{X}$ which satisfies $\mathcal{D} = \mathcal{D}^\perp$.

The notion of Dirac structures is suitable for the formulation of closed Hamiltonian systems, however, our aim is a treatment of open Hamiltonian systems in such a way that some of the external variables remain {\it free} port variables. For that reason, let $\mathcal{F}_b$ be a linear vector space of external flows, with the dual space $\mathcal{F}_b^\ast$ of external efforts. We deal with Dirac structures on the product space $\mathcal{X}\times \mathcal{F}_b$. The pairing on $(T\mathcal{X}\times \mathcal{F}_b )\oplus (T^\ast \mathcal{X} \times \mathcal{F}_b^\ast)$ is given by
\begin{equation}\label{eq:bilformgen}
\begin{split}
	&\left<\!\left< \Big((f_1,f_{b,1}), ( e_1,e_{b,1})\Big), \Big((f_2,f_{b,2}),  (e_2, e_{b,2})\Big) \right>\!\right> \\
	&\quad = 
	 \langle e_1|f_2\rangle + \langle e_{b,1}|f_{b,2}\rangle + \langle e_2|f_1\rangle+\langle e_{b,2}|f_{b,1}\rangle.
	\end{split}
\end{equation}
A \textbf{\emph{generalized Dirac structure}} $\mathcal{D}$ is a subbundle of $(T\mathcal{X}\times \mathcal{F}_b )\oplus (T^\ast \mathcal{X} \times \mathcal{F}_b^\ast)$ which is maximally isotropic under (\ref{eq:bilformgen}).\vspace{-0.2cm}

Consider a generalized Dirac structure $\mathcal{D}$ on the product space $\mathcal{X}\times \mathcal{F}_b$. Let $H: \mathcal{X}\rightarrow \mathbb{R}$ be a Hamiltonian. The \emph{\textbf{port-Hamiltonian system}} corresponding to a $4$-tuple $(\mathcal{X},\mathcal{F}_b,\mathcal{D},H)$ is defined by a set of smooth time-functions $\{t\mapsto (x(t),f_b(t),e_b(t))\in \mathcal{X}\times \mathcal{F}_b\times \mathcal{F}_b^\ast | t\in I\subset \mathbb{R})\}$ satisfying the equation
\begin{equation}\label{eq:finitedimpHsys}
\left( -\dot{x}(t),f_b(t), \mathrm{d} H(x(t)), e_b(t)\right)\in \mathcal{D} ~\; \mathrm{for}~\, t\in I.
\end{equation}
The equation (\ref{eq:finitedimpHsys}) implies the energy balance $
\frac{\mathrm{d} H}{\mathrm{d} t}(x(t))=\langle \mathrm{d} H (x(t))|\dot x(t)\rangle=\langle e_b(t)| f_b(t)\rangle$.

An important class of finite-dimensional port-Hamiltonian systems is given by
\begin{eqnarray}\label{eq:findimphsysJform}\notag
\dot{x}&=&J(x) \frac{\partial H}{\partial x}(x) + g(x) e_b\\
f_b &=& g^\textsc{t}(x) \frac{\partial H}{\partial x}\,,
\end{eqnarray}
where for clarity we have omitted the argument $t$, and $J:T^\ast \mathcal{X} \rightarrow T\mathcal{X}$ is a skew-symmetric vector bundle map and $g: \mathcal{F}_b\rightarrow T\mathcal{X}$ is the independent input vector field.

In this work, we deal exclusively with Dirac structures on linear spaces, which can be defined as follows. Let $\mathcal{F}$ and $\mathcal{E}$ be linear spaces. Given an $f\in\mathcal{F}$ and an $e\in\mathcal{E}$, the pairing will be denoted by $\langle e | f\rangle\in \mathbb{R}$. By symmetrizing the pairing, we obtain a symmetric bilinear form $\langle\!\langle,\rangle\!\rangle: \mathcal{F}\times \mathcal{E} \rightarrow \mathbb{R}$ naturally given as $
\langle\!\langle (f_1,e_1) ,(f_2,e_2)\rangle\!\rangle=\langle e_1 |f_2\rangle+\langle e_2 | f_1\rangle$.

A \textbf{\emph{constant Dirac structure}} is a linear subspace $\mathcal{D}\subset \mathcal{F} \times \mathcal{E}$ such that $\mathcal{D}=\mathcal{D}^{\perp}$, with $\perp $ standing for the orthogonal complement with respect to the bilinear form $\langle\!\langle,\rangle\!\rangle$.

\subsection{Stokes-Dirac structure}
The \emph{\textbf{Stokes-Dirac structure}} is an infinite-dimensional Dirac structure that provides a foundation for the port-Hamiltonian formulation of a class of distributed-parameter systems with boundary energy flow \cite{vdSM02}. 

Hereafter, let $M$ be an oriented $n$-dimensional smooth manifold with a smooth $(n-1)$-dimensional boundary $\partial M$ endowed with the induced orientation, representing the space of spatial variables. {{Adhering to the familiar ground in this paper, $M$ shall be a bounded Euclidian domain.}} By $\Omega^k(M)$, $k=0,1,\ldots,n$, denote the space of exterior $k$-forms on $M$, and by $\Omega^k(\partial M)$, $k=0,1,\ldots,n-1$, the space of $k$-forms on $\partial M$. 

For any pair $p,q$ of positive integers satisfying $p+q=n+1$, define the flow and effort linear spaces by
\begin{eqnarray*}
\mathcal{F}_{p,q}&=\,&\Omega^p(M)\times \Omega^q(M)\times \Omega^{n-p}(\partial M)\,\\
\mathcal{E}_{p,q}&=\,&\Omega^{n-p}(M)\times \Omega^{n-q}(M)\times \Omega^{n-q}(\partial M)\,.
\end{eqnarray*}

The bilinear form on the product space $\mathcal{F}_{p,q}\times\mathcal{E}_{p,q}$ is
\begin{eqnarray}\label{eq-aj9}\notag
\langle\!\langle & &\!\!\!\!\!\!\!(\underbrace{f_p^1, f_q^1, f_b^1}_{\in \mathcal{F}_{p,q}},\underbrace{e_p^1, e_q^1, e_b^1}_{\in \mathcal{E}_{p,q}}), (f_p^2, f_q^2, f_b^2, e_p^2, e_q^2, e_b^2)\rangle\!\rangle \\
& =&\int_M e_p^1 \wedge f_p^2+e_q^1\wedge f_q^2+ e_p^2\wedge f_p^1+e_q^2\wedge f_q^1\\
&&~ + \int_{\partial M} e_b^1 \wedge f_b^2 + e_b^2 \wedge f_b^1.\notag
\end{eqnarray}

\begin{thm}[\cite{vdSM02}]
Given linear spaces $\mathcal{F}_{p,q}$ and $\mathcal{E}_{p,q}$, and the bilinear form $\langle\!\langle, \rangle\!\rangle$, define the following linear subspace $\mathcal{D}$ of $\mathcal{F}_{p,q}\times \mathcal{E}_{p,q}$\vspace{-0.1cm}
\begin{eqnarray}\label{eq-7Cont}\notag
~~\mathcal{D}&=\big\{ &(f_p,f_q,f_b,e_p,e_q,e_b)\in \mathcal{F}_{p,q}\times \mathcal{E}_{p,q}\big |\\
& & \left[\begin{array}{c}f_p \\f_q\end{array}\right]=\left[\begin{array}{cc}0 & (-1)^{pq+1} {\mathrm{d}}\\ {\mathrm{d}} & 0\end{array}\right]\left[\begin{array}{c}e_p \\e_q\end{array}\right]\,,\\
& & \left[\begin{array}{c}f_b \\e_b\end{array}\right]=\left[\begin{array}{cc}\mathrm{tr}&0\\0 & -(-1)^{n-q}\mathrm{tr}\end{array}\right]\left[\begin{array}{c}e_p \\e_q\end{array}\right]
\big \}\,,\notag
\end{eqnarray}
where $\mathrm{d}$ is the exterior derivative and $\mathrm{tr}$ stands for the trace operator on the boundary $\partial M$. 
Then $\mathcal{D}=\mathcal{D}^\perp$, that is, $\mathcal{D}$ is a Dirac structure.
\end{thm}

Consider a Hamiltonian density $\mathcal{H}:\Omega^p(M) \times \Omega^q(M)\rightarrow \Omega^n(M)$ resulting in the Hamiltonian $H=\int_M \mathcal{H}\in \mathbb{ R}$. Setting the flows $f_p=-\frac{\partial \alpha_p}{\partial t}$, $f_q=-\frac{\partial \alpha_q}{\partial t}$ and the efforts $e_p=\delta_p H$, $e_q=\delta_q H$, where $(\delta_pH,\delta_qH)\in \Omega^{n-p}(M)\times \Omega^{n-q}(M)$ are the variational derivatives of $H$ at $(\alpha_p,\alpha_q)$, the \emph{\textbf{distributed-parameter port-Hamiltonian system}} is defined by the relation
\begin{equation}\label{eq:dpHsys}
\left(  -\frac{\partial \alpha_p}{\partial t},  -\frac{\partial \alpha_q}{\partial t},f_b, \delta_p H,\delta_qH, e_b\right)\in \mathcal{D}\,,~~t\in \mathbb{R}\,.\end{equation}
Since $\frac{\textmd{d} H}{\textmd{d} t}=\int_{\partial M}e_b\wedge f_b$, the system is \emph{lossless}.

\section{Basics of discrete exterior calculus}\label{Sec3}     

In the discrete setting, the smooth manifold $M$ is replaced by an oriented manifold-like simplicial complex. An $n$-dimensional \emph{\textbf{simplicial manifold}} $K$ is a simplicial triangulation of an $n$-dimensional polytope $|K|$ with an $(n-1)$-dimensional boundary. Familiar examples of such discrete manifolds are meshes of triangles embedded in $\mathbb{R}^3$ and tetrahedra obtained by tetrahedrization of $3$-dimensional manifolds.

\subsection{Chains and cochains}
The discrete analogue of a smooth $k$-form on the manifold $M$ is a $k$-cochain on the simplicial complex $K$. A $k$-chain is a formal sum of $k$-simplices of $K$ such that its value on a simplex changes sign when the simplex orientation is reversed. The free Abelian group generated by a basis consisting of oriented $k$-simplices with real-valued coefficients is $C_k(K;\mathbb{R})$. The space $C_k(K;\mathbb{R})$ is a vector space with dimension equal to the number of $k$-simplices in $K$, which is denoted by $N_k$. The space of $k$-cochains is the vector space dual of $C_k(K;\mathbb{R})$ denoted by $C^k(K;\mathbb{R})$ or $\Omega_d^k(K)$, as a reminder that this is the \emph{\textbf{space of discrete $k$-forms}}.



The \emph{\textbf{discrete exterior derivative}} or the \emph{\textbf{coboundary operator}} $\mathbf{d}^k:\Omega_d^k(K)\rightarrow \Omega_d^{k+1}(K)$ is defined by duality to the boundary operator $\partial_{k+1}:C_{k+1}(K;\mathbb{Z})\rightarrow C_{k}(K;\mathbb{Z})$, with respect to the natural pairing between discrete forms and chains. For a discrete form $\alpha\in \Omega_d^k(K) $ and a chain $c_{k+1}\in C_{k+1}(K;\mathbb{Z})$ we define $\mathbf{d}^k$ by
\begin{equation}\label{eq:discreteStokesTh}
\langle \mathbf{d}^k \alpha,c_{k+1}\rangle=\langle \alpha,(\mathbf{d}^k)^\textsc{t} c_{k+1}\rangle=\langle \alpha,\partial_{k+1}c_{k+1}\rangle\,,
\end{equation}
where the boundary operator $\partial_{k+1}$ is the \emph{incidence matrix} from the space of $(k+1)$-simplices to the space of $k$-simplices and is represented by a sparse $N_{k+1}\times N_{k}$ matrix containing only $0$ or $\pm1$ elements \cite{Desbrun2}. The important property of the boundary operator is $\partial_{k}\circ \partial_{k+1}=0$. The exterior derivative also satisfies $\mathbf{d}^{k+1}\circ\mathbf{d}^k =0$, what is a discrete analogue of the vector calculus identities $\mathrm{curl}\circ \mathrm{grad}=0$ and $\mathrm{div} \circ \mathrm{curl}=0$.

{{
\begin{rmk}
The relation (\ref{eq:discreteStokesTh}) can be regarded as a discrete {\textbf{{Stokes' theorem}}}, where the role of the exterior derivative is being played by the coboundary operator and the discrete analogue of integration is the evaluation of a cochain.
\end{rmk}
}}


{{
\subsection{Dual cell complex}
An essential ingredient of discrete exterior calculus is the \textbf{\emph{dual complex}} of a manifold-like simplicial complex. The main idea is to associate to each {\it primal} $k$-simplex a {\it dual} $(n-k)$-cell. For example, in the $3$-dimensional case, consider a tetrahedral mesh with interior elements shown in Fig.~\ref{fig:3dex}. We associate a dual $3$-cell to each primal vertex ($0$-simplex), a dual polygon ($2$-cell) to each primal edge ($1$-simplex), a dual edge ($1$-cell) to each primal face ($2$-simplex), and a dual vertex ($0$-cell) to each primal tetrahedron ($3$-simplex).

\begin{figure}
\centering
\vspace{-0.0cm}
      \hspace{0cm}\includegraphics[width=8.2cm]{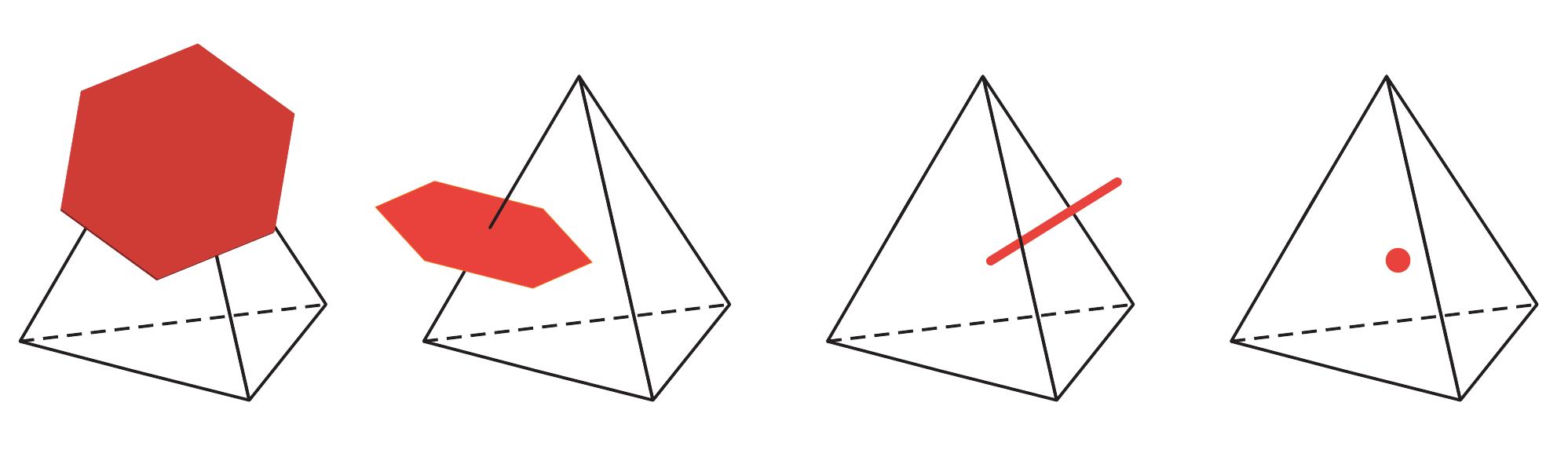}
      \vspace{-0.1cm}
  \caption{A $3$-dimensional example of primal and dual mesh elements. The corresponding circumcentric dual cells are shaded.}\label{fig:3dex}
\end{figure}
\vspace{-0.2cm}

In the $2$-dimensional case, for illustration consider the triangular mesh in Fig.~\ref{fig:RDcomp12}. To the primal edge $[v_i,v_j]$ we associate the dual edge $[\hat{v}_i,\hat{v}_j]$, where the vertices $\hat{v}_i$ and $\hat{v}_j$ are the circumcenters\footnote{The circumcenter of a $k$-simplex $\sigma^k$ is given by the centre of the $k$-circumsphere, which is the unique $k$-sphere that has all $k+1$ vertices of $\sigma^k$ on its surface.} of the two neighbouring triangles that share the common edge $[v_i,v_j]$. The dual edge of $[v_i,v_j]$ will be denoted by $\star_{\mathrm{i}}[v_i,v_j]$. The dual of the vertex $v_r$ is its Voronoi region shown shaded. The dual of the face $[v_m,v_p,v_n]$ is its circumcenter $\hat{v}_r$, while the dual of the edge $[v_k,v_l]$ is the (half-)edge $[\hat{v}_k, \hat{v}_l]=\star_{\mathrm{i}}[v_k,v_l]$ orthogonal to $[v_k,v_l]$ and restricted to $|K|$.

\begin{figure}
   \hspace{-0.54cm}\includegraphics[width=9.1cm]{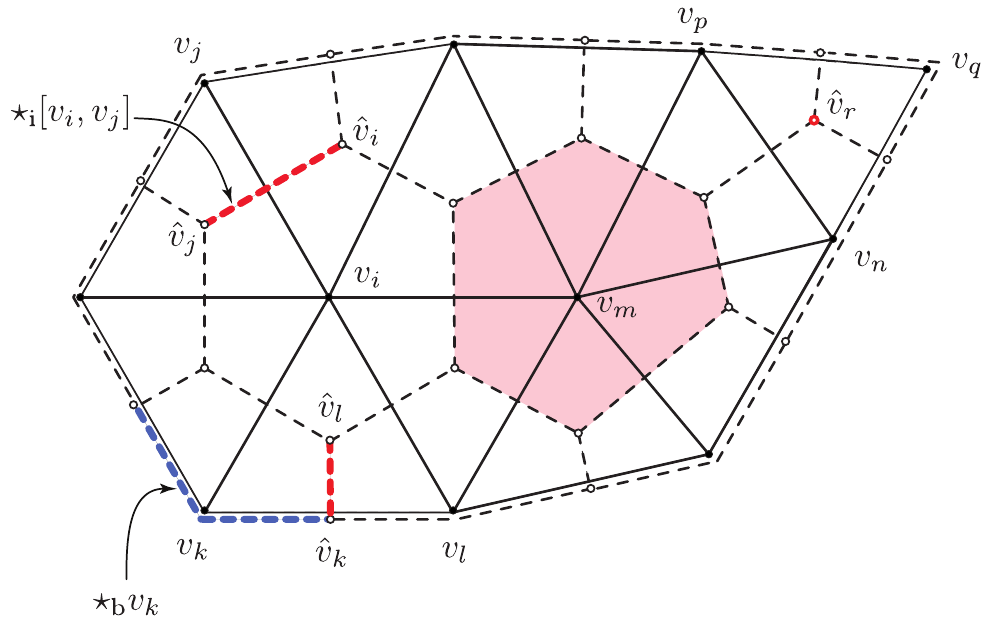}\\\vspace{-0.4cm}
  \caption{A $2$-dimensional simplicial complex $K$ and its circumcentric dual cell complex $\star K$ indicated by dashed lines. The boundary of $\star K$ is the dual of the boundary of $K$.}\label{fig:RDcomp12}
\end{figure}

The just explained geometric duality is the so-called {\it circumcentric} or Voronoi duality\footnote{In algebraic topology \cite{Munkres} and computational electromagnetics \cite{Bosavit,Hiptmair}, another popular choice of the geometric dualism is barycentric duality.}, which has an important property that primal and dual cells are orthogonal to each other. This feature dramatically simplifies the discrete counterpart of the Hodge star, as will be shown in the next subsection. For this reason we shall be dealing with the circumcentric duality and require that the simplicial complex is well-centred (the circumcenters of all simplices of all dimensions lie in the interior of the corresponding simplices).

Given a simplicial  complex $K$, we define its \emph{interior dual cell complex} $\star_\mathrm{i} K$ as the circumcentric dual of $K$ geometrically restricted to $|K|$.

In a similar fashion, to each primal $k$-simplex living on the geometric boundary of $K$, hereafter denoted by $\partial K$, we can uniquely associate an $(n-1-k)$-cell living on the dual of the boundary $\partial K$. The circumcentric dual of $\partial K$ is the \emph{boundary dual cell complex} $\star_\mathrm{b} K$. For example, considering Fig.~\ref{fig:RDcomp12}, on the boundary, the dual of the edge $[v_k,v_l]$ is the dual vertex $\hat{v}_k$, while the boundary dual of the primal vertex $v_k$ is the curvilinear edge $\star_{\mathrm{b}}v_k$ shown bolded.

The dual cell complex $\star K$ is defined as $\star K=\star_\mathrm{i} K \times \star_\mathrm{b} K$. The dual mesh $\star_\mathrm{i} K$ is a dual to $K$ in sense of a graph dual, and the dual of the boundary is equal to the boundary of the dual, that is $\partial (\star K)=\star (\partial K)=\star_\mathrm{b} K$. Because of duality, there is: (1) a one-to-one correspondence between $k$-simplices of $K$ and interior $(n-k)$-cells of $\star K$; (2) a bijection between primal $k$-simplices of $\partial K$ and the dual boundary $(n-1-k)$-cells of $\star_\mathrm{b}K$.
}}

\subsection{Exterior derivatives on the dual mesh}\vspace{-0.2cm}
Everything that has been said about the primal discrete forms carries over to the dual cochains, which can be interpreted as covectors. The space of dual $k$-cochains will be denoted as $\Omega_d^{k}(\star_\mathrm{i} K)$. The covectors will be labeled by a caret symbol, e.g., $\hat\beta\in\Omega_d^{k}(\star_\mathrm{i} K) $.

The \emph{\textbf{trace operator}} $\mathbf{tr}^k:\Omega_d^k(K)\rightarrow\Omega_d^k(\partial K)$ is a matrix that isolates the members of a $k$-cochain vector assumed on the geometric boundary $\partial K$.

The \emph{\textbf{dual exterior derivative}} $\mathbf{d}_\mathrm{i}^{n-k}:\Omega_d^{n-k}(\star_\mathrm{i} K)\rightarrow \Omega_d^{n-k+1}(\star_\mathrm{i} K)$ is defined by duality to the primal exterior operator $\mathbf{d}^k$ as
\begin{equation*}
\mathbf{d}_\mathrm{i}^{n-k}=(-1)^k (\mathbf{d}^{k-1})^\textsc{t}\,.
\end{equation*}
The negative sign appears as the orientation of the dual is induced by the primal orientation. 

The \emph{\textbf{dual boundary exterior derivative}} $\mathbf{d}_\mathrm{b}^{n-k}:\Omega_d^{n-k}(\star_\mathrm{b} K)\rightarrow \Omega_d^{n-k+1}(\star_\mathrm{i} K)$ is defined as
\[
\mathbf{d}_\mathrm{b}^{n-k}=(-1)^{k-1} (\mathbf{tr}^{k-1})^\textsc{t}\,.
\]


\begin{figure}
\centering
\vspace{-0.0cm}
      \hspace{0cm}\includegraphics[width=4.6cm]{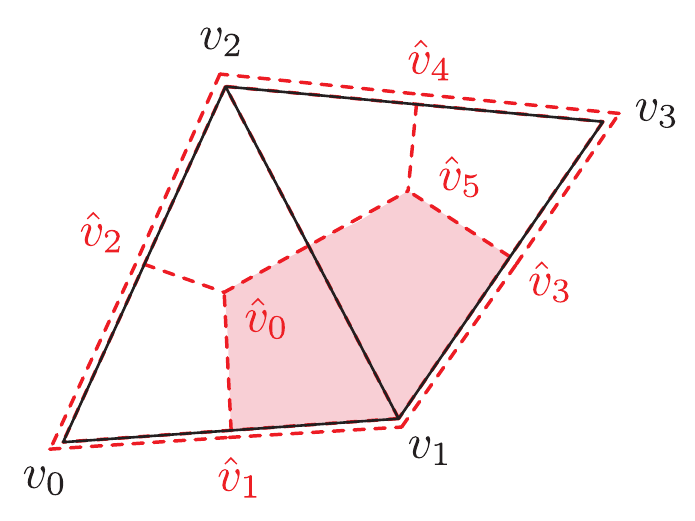}
      \vspace{-0.1cm}
  \caption{The simplicial complex $K$ consisting of two triangles. The dual edges introduced by subdivision are shown dotted. The shaded area is the dual cell $\star_\mathrm{i} v_1$ of the primal vertex $v_1$.}\label{fig:wave}
\end{figure}

{{
\vspace{-0.2cm}
\begin{exm}\label{Example1}
Consider a simplicial complex pictorially given by Fig.~\ref{fig:wave}. The primal and dual $2$-faces have counterclockwise orientations. The matrix representation of the incidence operator $\partial_1$, from the primal edges to the primal vertices, is
\begin{equation*}\vspace{-0.2cm}
\begin{array}{cccccc}
~\, & [v_0,v_1] & [v_1,v_2] & [v_2,v_0] & [v_1,v_3] & [v_3,v_2] \\
v_0 & -1 & ~0 & ~0 & ~0 & ~0 \\
v_1 & ~1 & -1 & ~0 & -1 & ~0 \\
v_2 & ~0 & ~1 & -1 & ~0 & ~1 \\
v_3 & ~0 & ~0 & ~1 & ~1 & -1\end{array}
\end{equation*}
while the discrete exterior derivative from the vertices to the edges is the transpose of the incidence operator, i.e., $\mathbf{d}^0=\partial_1^\textsc{t}$. The dual exterior derivative is $\mathbf{d}_\mathrm{i}^1=-\left(\mathbf{d}^0\right)^\textsc{t}$, while the matrix representation of the $\mathbf{d}_\mathrm{b}^1$ operator is\vspace{-0.2cm}
\begin{equation*}
\begin{array}{lcccc}
~ & \left[\hat{v}_2,\hat{v}_1\right] & \left[\hat{v}_1,\hat{v}_3\right]   & \left[\hat{v}_3,\hat{v}_4\right]   & \left[\hat{v}_4,\hat{v}_2\right]  \\
\star_{\mathrm{i}} v_0 & 1 & 0 & 0 & 0 \\
\star_{\mathrm{i}} v_1 & 0 & 1 & 0 & 0  \\
\star_{\mathrm{i}} v_2 & 0 & 0 & 0 & 1 \\
\star_{\mathrm{i}} v_3& 0 & 0 & 1 & 0\end{array}
\end{equation*}Here, as previously explained, $\star_{\mathrm{i}} v_j$ is the Voronoi dual of a vertex $v_j$, and $\left[\hat{v}_2,\hat{v}_1\right], \left[\hat{v}_1,\hat{v}_3\right]   , \left[\hat{v}_3,\hat{v}_4\right]   , \left[\hat{v}_4,\hat{v}_2\right]$ are the boundary duals of $v_0,v_1,v_2,v_3$, respectively.

The trace operator is $\mathbf{tr}^0=  ( \mathbf{d}_\mathrm{b}^1 )^\textsc{t}$.

The incidence operator $\partial_2$, from the set of primal faces to the set of the primal edges, is
\begin{equation*}\vspace{-0.3cm}
\begin{array}{lccc}
~ & \left[v_0,v_1,v_2\right] & \left[v_1,v_3,v_2\right] \\
 \left[v_0,v_1\right] & 1 & 0  \\
 \left[v_1,v_2\right] & 1 & -1 \\
 \left[v_2,v_0\right] & 1 & 0 \\
 \left[v_1,v_3\right]& 0 & 1 \\
 \left[v_3,v_2\right]& 0 & 1 \\
 \end{array}
\end{equation*}

The $\mathbf{d}_\mathrm{b}^0$ operator is\vspace{-0.3cm}
\begin{equation*}
\begin{array}{lcccc}
~ & ~~~\hat{v}_1~~~ & ~~~\hat{v}_2~~~   & ~~~\hat{v}_3~~~   & ~~~\hat{v}_4 ~~~ \\
\left[\hat{v}_1,\hat{v}_0\right] & -1 & 0 & 0 & 0 \\
\left[\hat{v}_5,\hat{v}_0\right] & 0 & 0 & 0 & 0  \\
\left[\hat{v}_2,\hat{v}_0\right] & 0 & -1 & 0 & 0 \\
\left[\hat{v}_3,\hat{v}_5\right] & 0 & 0 & -1 & 0\\
\left[\hat{v}_4,\hat{v}_5\right] & 0 & 0 & 0 & -1
\end{array}
\end{equation*}
The trace operator that isolates the elements living on the boundary edges is $\mathbf{tr}^1=  -( \mathbf{d}_\mathrm{b}^0 )^\textsc{t}$.
\end{exm}
}}

\subsection{Discrete wedge and Hodge operator}
There exists a natural pairing, via the so-called primal-dual wedge product, between a primal $k$-cochain and a dual $(n-k)$-cochain. Let $\alpha\in \Omega_d^k(K)$ and $\hat{\beta}\in \Omega_d^{n-k}(\star_\mathrm{i} K)$. {{We define the discrete \emph{\textbf{primal-dual wedge product}} $\wedge: \Omega_d^k(K)\times \Omega_d^{n-k}(\star_\mathrm{i} K) \rightarrow \mathbb{R}$ by\vspace{-0.2cm}
\begin{eqnarray*}
\langle \alpha\wedge \hat{\beta},K\rangle
&=&\sum_{\sigma_j^k\in K}\langle \alpha,\sigma_j^k \rangle \langle \hat{\beta}, \star_\mathrm{i}\sigma_j^k\rangle =\alpha^\textsc{t} \hat{\beta}\\
&=&(-1)^{k(n-k)}\langle \hat{\beta}\wedge \alpha,K\rangle\,,\vspace{-0.2cm}
\end{eqnarray*}
where the summation is taken over all $k$-simplices $\sigma_j^k$ in $K$. Analogously, for an $\alpha_\mathrm{b}\in\Omega_d^k(\partial K)$ and a $\hat{\beta}_\mathrm{b}\in\Omega_d^{n-k-1}(\star_\mathrm{b}K)$, we define the primal-dual pairing on the boundary by
\begin{eqnarray*}
\langle \alpha_\mathrm{b}\wedge \hat{\beta}_\mathrm{b},\partial K\rangle
&=&\alpha_\mathrm{b}^\textsc{t} \hat{\beta}_\mathrm{b}=(-1)^{k(n-k-1)}\langle \hat{\beta}_\mathrm{b}\wedge \alpha_\mathrm{b},\partial K\rangle\,.
\end{eqnarray*}


\begin{rmk}
Given a primal $(k-1)$-form $\alpha$ and an internal dual $(n-k)$-discrete form $\hat{\beta}_\mathrm{i} \in \Omega_d^{n-k}(\star_\mathrm{i}K)$ and a dual boundary form $\hat{\beta}_\mathrm{b}\in \Omega_d^{n-k}(\star_\mathrm{b}K) $, then
\begin{eqnarray*}
\langle \mathbf{d}^{k-1} \alpha \wedge \hat{\beta}_\mathrm{i} ,K\rangle\!&+&\!(-1)^{k-1}\!\langle \alpha \wedge ( \mathbf{d}_\mathrm{i}^{n-k} \hat{\beta}_\mathrm{i} +\mathbf{d}_\mathrm{b}^{n-k} \hat{\beta}_\mathrm{b} ),K\rangle \\
&&~~~~~~~~~~~~~~~~~~~=\langle  \mathbf{tr}^{k-1}\alpha \wedge \hat{\beta}_\mathrm{b} ,\partial K\rangle\,.
\end{eqnarray*}
The last relation is the {\it{\textbf{summation-by-parts formula}}} and its validity is an immediate consequence of $\mathbf{d}_\mathrm{i}^{n-k}=(-1)^k (\mathbf{d}^{k-1})^\textsc{t}$ and $\mathbf{d}_\mathrm{b}^{n-k}=(-1)^{k-1} (\mathbf{tr}^{k-1})^\textsc{t}$.
\end{rmk}

The support volumes of a simplex and its dual cell are the same, which suggests that there is a natural identification between primal $k$-cochains and dual $(n-k)$-cochains. In the exterior calculus for smooth manifolds, the Hodge star, denoted $*_k$, is an isomorphism between the space of $k$-forms and $(n-k)$-forms. The \emph{\textbf{discrete Hodge star}} is a map $*_k : \Omega_d^k(K)\rightarrow \Omega_d^{n-k}(\star_\mathrm{i} K)$ defined by its value over simplices and their duals. In case of the circumcentric duality, the Hodge star $*_k$ is a diagonal $N_k\times N_k$ matrix with the entry corresponding to a simplex $\sigma^k$ being $|\sigma^k|/|\star_\mathrm{i} \sigma^k|$, that is
\[
*_k=\mathrm{diag}\left( \frac{|\star_\mathrm{i} \sigma_1^k|}{|\sigma_1^k|},\frac{|\star_\mathrm{i} \sigma_2^k|}{|\sigma_2^k|},\cdots, \frac{|\star_\mathrm{i} \sigma_{N_k}^k|}{|\sigma_{N_k}^k|} \right).
\]
Here, $|\sigma^k|$ and $|\star_\mathrm{i} \sigma^k|$ are the volumes of $\sigma^k$ and $\star_\mathrm{i} \sigma^k$, respectively\footnote{The convention is that $|\sigma^0|=1$.}. 

\begin{exm}Consider the $2$-dimensional simplicial complex and its circumcentric dual in Fig.~\ref{fig:wave}. The diagonal Hodge operators are
\begin{eqnarray*}
*_0 &=& \mathrm{diag}\left(  {|\star_\mathrm{i} v_0|} , {|\star_\mathrm{i} v_1|} , {| \star_\mathrm{i}v_2|}, {|\star_\mathrm{i} v_3|}  \right)\\
*_1 &=& \mathrm{diag}\bigg( \frac{|[\hat v_1, \hat v_0]|}{|[v_0,v_1]|}, \frac{|[\hat v_0, \hat v_5]|}{|[v_1,v_2]|}, \frac{|[\hat v_2, \hat v_0]|}{|[v_2,v_0]|}, \frac{|[\hat v_3, \hat v_5]|}{|[v_1,v_3]|}, \\&&~~~~~~~~~~~~~~~~~~~~~~~~~~~~~~~~~~~~~~~~~~~~~~~~~~\frac{|[\hat v_4, \hat v_5]|}{|[v_3,v_2]|} \bigg)\\
*_2 &=& \mathrm{diag}\bigg( \frac{1}{|[v_0,v_1,v_2]|},  \frac{1}{|[v_1,v_3,v_2]|}\bigg)\,.
\end{eqnarray*}
\end{exm}\vspace{-0.5cm}
}}

\begin{rmk}Another possibility for the construction of the Hodge operator is to use Whitney forms. The Whitney map is an interpolation scheme for cochains. It maps discrete forms to square integrable forms that are piecewise smooth on each simplex. The Whitney maps are built from barycentric coordinate functions and the resulting matrix is sparse but in general \emph{not} diagonal \cite{Bosavit}, \cite{Hiptmair}.
\end{rmk}

The linear operators of discrete exterior calculus used in this paper are succinctly presented in the following diagram\vspace{-0.99cm}\small
\begin{equation*} \label{prescription}
\xymatrixcolsep{1.8pc}\xymatrixrowsep{2.6pc}\xymatrix{
\Omega_d^0(\partial K)& \ar[l]^{\mathbf{tr}^0}\Omega_d^0(K)\ar[d]^{\mathbf{d}^0} \ar@<0.4ex>[r]^{*_0}  & \ar@<0.5ex>[l]^{*_0^{-1}}\Omega_d^n(\star_\mathrm{i} K) &\ar[l]_{\mathbf{d}_\mathrm{b}^{n-1}}\Omega_d^{n-1}\! (\star_\mathrm{b} K) \\
\Omega_d^1(\partial K)& \ar[l]^{\mathbf{tr}^1}\Omega_d^1(K)\ar[d]^{\mathbf{d}^1}\ar@<0.4ex>[r]^{*_1~~} & \ar@<0.5ex>[l]^{*_1^{-1}} \Omega_d^{n-1}(\star_\mathrm{i} K)\ar[u]_{\mathbf{d}_\mathrm{i}^{n-1}} &\ar[l]_{~~~\mathbf{d}_\mathrm{b}^{n-2}}\Omega_d^{n-2} (\star_\mathrm{b} K)\\
\vdots& \vdots \ar[d]^{\mathbf{d}^{n-2}} & \vdots \ar[u]_{\mathbf{d}_\mathrm{i}^{n-2}} & \vdots \\
\!\Omega_d^{n-1}(\partial K)& \ar[l]^{~~~\mathbf{tr}^{n-1}}\Omega_d^{n-1}(K)\ar[d]^{\mathbf{d}^{n-1}} \ar@<0.4ex>[r]^{*_{n-1}} &  \ar@<0.4ex>[l]^{*_{n-1}^{-1}}\Omega_d^{1}(\star_\mathrm{i} K)\ar[u]_{\mathbf{d}_\mathrm{i}^{1}} &\ar[l]_{\mathbf{d}_\mathrm{b}^{0}}\Omega_d^{0} (\star_\mathrm{b} K)\\
~&  \Omega_d^{n}(K) \ar@<0.4ex>[r]^{*_{n}~~} & \ar@<0.4ex>[l]^{*_{n}^{-1}}\Omega_d^{0}(\star_\mathrm{i} K)\ar[u]_{\mathbf{d}_\mathrm{i}^{0}} & ~\,
}
\end{equation*}\normalsize




\section{Simplicial Dirac structures}\label{Sec4}
In this section, we develop the matrix representations of \emph{\textbf{simplicial Dirac structures}}. These structures are discrete analogues of the Stokes-Dirac structure and as such are defined in terms of primal and duals cochains on the underlying discrete manifold.\vspace{-0.3cm}

In the discrete setting, the role of the bounded domain $M$ is played by an $n$-dimensional well-centred oriented manifold-like simplicial complex $K$. The flow and the effort spaces will be the spaces of complementary primal and dual forms. The elements of these two spaces are paired via the discrete primal-dual wedge product. Let
\begin{eqnarray*}
\mathcal{F}_{p,q}^d&=&\Omega_d^p(\star_\mathrm{i} K)\times \Omega_d^q( K)\times \Omega_d^{n-p}(\partial (K))\\
\mathcal{E}_{p,q}^d&=&\Omega_d^{n-p}( K)\times \Omega_d^{n-q}(\star_\mathrm{i} K)\times \Omega_d^{n-q}(\partial (\star K))\,.
\end{eqnarray*}
The primal-dual wedge product ensures a \emph{bijective} relation between the primal and dual forms, between the flows and efforts. A natural discrete mirror of the bilinear form (\ref{eq-aj9}) is a symmetric pairing on the product space $\mathcal{F}_{p,q}^d\times \mathcal{E}_{p,q}^d$ defined by\vspace{-0.3cm}
\begin{eqnarray}\label{eq:bild}\notag
\langle\!\langle (&&\!\!\!\!\!\!\underbrace{\hat{f}_p^1,{f}_q^1,{f}_b^1}_{\in \mathcal{F}_{p,q}^d},\underbrace{{e}_p^1,\hat{e}_q^1,\hat{e}_b^1}_{\in \mathcal{E}_{p,q}^d}), (\hat{f}_p^2,{f}_q^2,{f}_b^2,{e}_p^2,\hat{e}_q^2,\hat{e}_b^2)\rangle\!\rangle_d\\
&=& \langle {e}_p^1\wedge \hat{f}_p^2+\hat{e}_q^1\wedge {f}_q^2+ {e}_p^2\wedge \hat{f}_p^1+\hat{e}_q^2\wedge {f}_q^1,K \rangle\\\notag
&&~\;~+\langle \hat{e}_b^1\wedge {f}_b^2+ \hat{e}_b^2\wedge f_b^1,\partial K \rangle
  \,.
\end{eqnarray}
A discrete analogue of the Stokes-Dirac structure is the finite-dimensional Dirac structure constructed in the following theorem.

\vspace{-0.2cm}
\begin{thm}\label{th:pdDirac}
Given linear spaces $\mathcal{F}_{p,q}^d$ and $\mathcal{E}_{p,q}^d$, and the bilinear form $\langle\!\langle, \rangle\!\rangle_d$. The linear subspace $\mathcal{D}_d\subset\mathcal{F}_{p,q}^d\times \mathcal{E}_{p,q}^d$ defined by
\begin{eqnarray}\label{eq:Dir-prim-dual}\notag
& & \mathcal{D}_{d}=\big\{ (\hat{f}_p, {f}_q, {f}_b,{e}_p,\hat{e}_q,\hat{e}_b)\in \mathcal{F}_{p,q}^d\times \mathcal{E}_{p,q}^d\big |\\\notag
\!\!\!\!\!& & \left[\begin{array}{c}\hat{f}_p \\ {f}_q \end{array} \right] \!\!=\!\!\left[\begin{array}{cc} 0  & (-1)^{r}  \mathbf{d}_\mathrm{i}^{n-q}\\  \mathbf{d}^{n-p} & 0\end{array} \right] \left[\begin{array}{c} {e}_p \\ \hat{e}_q\end{array} \right]+ (-1)^{r} \left[\begin{array}{c}  \mathbf{d}_\mathrm{b}^{n-q} \\ 0 \end{array}\right] \hat{e}_b\,,\\
& &\begin{array}{c}~~~~f_b\, \end{array}\!\!\!=\!\! ~(-1)^{p}\mathbf{tr}^{n-p}{e}_p \}\,,
\end{eqnarray}
with $r=pq+1$, is a Dirac structure with respect to the pairing $\langle\!\langle, \rangle\!\rangle_{d}$ .
\end{thm}
\vspace{-0.55cm}
\begin{pf}Note that since $\mathbf{d}_\mathrm{i}^{n-q}=(-1)^q (\mathbf{d}^{n-p})^\textsc{t}$ and $\mathbf{d}_\mathrm{b}^{n-q}=(-1)^{n-p}(\mathbf{tr}^{n-p})^\textsc{t}$, the operator
\begin{equation*}
\left[\begin{array}{ccc} 0  & (-1)^{r}  \mathbf{d}_\mathrm{i}^{n-q} &(-1)^{r} \mathbf{d}_\mathrm{b}^{n-q}   \\  \mathbf{d}^{n-p} & 0 & 0\\
(-1)^{p}\mathbf{tr}^{n-p} & 0 &
\end{array} \right]
\end{equation*}
is skew-symmetric, and thus (\ref{eq:Dir-prim-dual}) is a \emph{Poisson structure} on the state space $\Omega_d^p( \star_\mathrm{i}K)\times \Omega_d^q( K)$.\qed
\end{pf}



The other discrete analogue of the Stokes-Dirac structure is defined on the spaces
\begin{eqnarray*}
\tilde{\mathcal{F}}_{p,q}^d&=&\Omega_d^p (K)\times \Omega_d^q( \star_\mathrm{i} K)\times \Omega_d^{n-p}(\partial (\star K))\\
\tilde{\mathcal{E}}_{p,q}^d&=&\Omega_d^{n-p}( \star_\mathrm{i} K)\times \Omega_d^{n-q}( K)\times \Omega_d^{n-q}(\partial K)\,.
\end{eqnarray*}
A natural discrete mirror of (\ref{eq-aj9}) in this case is a symmetric pairing defined by
\begin{eqnarray*}
\langle\!\langle (&\!\!\underbrace{{f}_p^1,\hat{f}_q^1,{\hat{f}}_b^1}_{\in \tilde{\mathcal{F}}_{p,q}^d},\underbrace{{\hat{e}}_p^1,{e}_q^1,{e}_b^1}_{\in \tilde{\mathcal{E}}_{p,q}^d}), ({f}_p^2,\hat{f}_q^2,\hat{f}_b^2,\hat{e}_p^2,{e}_q^2,{e}_b^2)\rangle\!\rangle_{\tilde{d}} \\
&= \langle \hat{e}_p^1\wedge {f}_p^2+ {e}_q^1\wedge {\hat{f}}_q^2+ {\hat{e}}_p^2\wedge {f}_p^1+ {e}_q^2\wedge \hat{f}_q^1,K \rangle\\
&\!\!\!\!\!\!\!\!\!\!\!\!\!\!\!\!\!+\langle {e}_b^1\wedge \hat{f}_b^2+ {e}_b^2\wedge \hat{f}_b^1,\partial K \rangle\,.
\end{eqnarray*}

\vspace{-0.2cm}
\begin{thm}\label{th:pdDiracN}
The linear space $\tilde{\mathcal{D}}_d$ defined by 
\begin{eqnarray}\label{eq:Dir-prim-dual2}\notag
&&\tilde{\mathcal{D}}_d=\big\{ ({f}_p, \hat{f}_q, {f}_b,{e}_p, {e}_q, {e}_b)\in\tilde{ \mathcal{F}}_{p,q}^d\times \tilde{\mathcal{E}}_{p,q}^d\big |\\\notag
&& \left[\!\!\begin{array}{c} {f}_p \\ {f}_q\end{array}\!\!\right]=\left[\!\!\begin{array}{cc}0 & (-1)^{pq+1}  \mathbf{d}^{n-q} \\  \mathbf{d}_\mathrm{i}^{n-p} & 0\end{array}\!\!\right]\left[\!\!\begin{array}{c} \hat{e}_p \\ {e}_q\end{array}\!\!\right]+\left[\!\!\begin{array}{c}  0\\ \mathbf{d}_\mathrm{b}^{n-p} \end{array}\!\!\right] \hat{f}_b\,,\\
&&\begin{array}{c}\,\,~e_b ~\end{array}= ~(-1)^{p}\mathbf{tr}^{n-q}{e}_q\,
\big \}\,
\end{eqnarray}
is a Dirac structure with respect to the bilinear pairing $\langle\!\langle, \rangle\!\rangle_{\tilde{d}}$.
\end{thm}
\vspace{-0.4cm}
\begin{pf}
The simplicial Dirac structure (\ref{eq:Dir-prim-dual2}) is the dual of (\ref{eq:Dir-prim-dual}), and the proof is analogue to that of Theorem~\ref{th:pdDirac}.\qed
\end{pf}\vspace{-0.6cm}
In the following section, the simplicial Dirac structures (\ref{eq:Dir-prim-dual}) and (\ref{eq:Dir-prim-dual2}) will be used as \emph{terminus a quo} for the geometric formulation of spatially discrete port-Hamiltonian systems.

\section{Port-Hamiltonian systems}\label{Sec5}
Let a function $\mathcal{H}: \Omega_d^p(\star_\mathrm{i} K)\times \Omega_d^q(K)\rightarrow \mathbb{R}$ stand for the Hamiltonian 
$(\hat{\alpha}_p,\alpha_q)\mapsto \mathcal{H}(\hat{\alpha}_p,\alpha_q)$, with $\hat{\alpha}_p \in \Omega_d^p(\star_\mathrm{i} K)$ and $\alpha_q\in \Omega_d^q(K)$. A time derivative of $\mathcal{H}$ along an arbitrary trajectory $t\rightarrow (\hat{\alpha}_p(t),\alpha_q(t))\in  \Omega_d^p(\star_\mathrm{i} K)\times  \Omega_d^q( K)$, $t\in \mathbb{R}$, is\vspace{-0.2cm}
\begin{eqnarray}
\frac{\mathrm{d}}{\mathrm{d}t}\mathcal{H}(\hat{\alpha}_p,\alpha_q)=\left\langle \frac{\partial \mathcal{H}}{\partial \hat{\alpha}_p} \wedge \frac{\partial \hat{\alpha}_p}{\partial t} + \hat{\frac{\partial \mathcal{H}}{\partial \alpha_q}}\wedge \frac{\partial \alpha_q }{\partial t},K\right\rangle\,,
\end{eqnarray}
where the caret sign reminds that the quantity lives on the dual mesh. The relations between the simplicial-Dirac structure (\ref{eq:Dir-prim-dual}) and time derivatives of the variables are: $\hat{f}_p=-\frac{\partial \hat{\alpha}_p}{\partial t}$, $f_q=-\frac{\partial \alpha_q }{\partial t}$, while the efforts are: $e_p=\frac{\partial \mathcal{H}}{\partial \hat{\alpha}_p}$, $\hat{e}_q=\hat{\frac{\partial \mathcal{H} }{\partial \alpha_q}}$.

This allows us to define a time-continuous port-Hamiltonian system on a simplicial complex $K$ (and its dual $\star K$) by
\begin{eqnarray}\label{eq-38d}\notag
\!\!\!\!\!\!\!\!\!\!\!\!\!\!\!\!\!\!\!\!\!\!\left[\!\!\begin{array}{c}-\frac{\partial\hat{\alpha}_p}{\partial t} \\ -{\frac{\partial\alpha_q}{\partial t}}\end{array}\!\!\!\right]\!\!&=&\!\!\left[\!\!\begin{array}{cc}0 & \!\!\!\!(-1)^{r}  \mathbf{d}_\mathrm{i}^{n-q}\\  \mathbf{d}^{n-p} & 0\end{array}\!\!\right]\!\!\left[\!\!\begin{array}{c} {\frac{\partial \mathcal{H}}{\partial \hat{\alpha}_p}} \\\hat{\frac{\partial \mathcal{H}}{\partial \alpha_q}}\end{array}\!\!\right]+\!(-1)^{r}\!\!\left[\!\!\begin{array}{c}  \mathbf{d}_\mathrm{b}^{n-q} \\ 0 \end{array}\!\!\right] \hat{e}_b\,,\!\!\!\!\\
\begin{array}{c}~~\,\,f_b ~~\end{array}&=& ~(-1)^{p}\mathbf{tr}^{n-p}\frac{\partial \mathcal{H}}{\partial \hat{\alpha}_p}\,,
\end{eqnarray}
where $r=pq+1$.

The system (\ref{eq-38d}) is evidently in the form (\ref{eq:findimphsysJform}). It immediately follows that $\frac{\mathrm{d}}{\mathrm{d}t}\mathcal{H}= \langle \hat{e}_b \wedge {f}_b, \partial K \rangle $, enunciating a fundamental property of the system: the increase in the energy on the domain $|K|$ is equal to the power supplied to the system through the boundary $\partial K$ and $\partial (\star K)$. The boundary efforts $\hat{e}_b$ are the boundary control input and $f_b$ are the outputs. 

\vspace{0.0cm}
\begin{rmk}\label{rmk:pass}
Introducing a linear negative feedback control as $\hat{e}_b=(-1)^{(n-p)(n-q)-1}*_\mathrm{b}f_b$, where $*_\mathrm{b}$ is the Hodge star on the boundary $\partial K$, leads to passivization of the lossless port-Hamiltonian system, i.e., $\frac{\mathrm{d}}{\mathrm{d}t}\mathcal{H}\leq-  \langle {f}_b \wedge *_\mathrm{b} {f}_b, \partial K \rangle \leq 0$. Furthermore, if the Hamiltonian is a $\mathcal{K}_\infty$ function with a strict minimum that is a stationary set for the system (\ref{eq-38d}), the equilibrium is asymptotically stable. A more elaborate control strategy can be the energy shaping method as is briefly discussed in Section~\ref{Sec:ECmethod}.
\end{rmk}

An alternative formulation of a spatially discrete port-Hamiltonian system is given in terms of the simplicial Dirac structure (\ref{eq:Dir-prim-dual2}). We start with the Hamiltonian function $
({\alpha}_p, \hat{\alpha}_q)\mapsto \mathcal{H}({\alpha}_p, \hat{\alpha}_q)$, where ${\alpha}_p \in \Omega_d^p(K)$ and $\hat{\alpha}_q\in \Omega_d^q(\star _\mathrm{i}K)$.
In a similar manner as in deriving (\ref{eq-38d}), we introduce the input-output port-Hamiltonian system
\begin{eqnarray}\label{eq-38dN}\notag
\left[\!\begin{array}{c}\!-\frac{\partial{\alpha}_p}{\partial t} \\ \!\!-{\frac{\partial \hat{\alpha}_q}{\partial t}}\!\!\!\end{array}\right]&\!=&\!\left[\!\!\begin{array}{cc}0 & \!\!\!(-1)^{r}  \mathbf{d}^{n-q}\\  \mathbf{d}_\mathrm{i}^{n-p} & 0\end{array}\!\!\right]\!\left[\!\!\begin{array}{c} \hat{{\frac{\partial \mathcal{H}}{\partial{\alpha}_p}}} \\ {\frac{\partial \mathcal{H}}{\partial \hat{\alpha}_q}}\end{array}\!\!\right]\!+\!\left[\!\!\begin{array}{c} 0\\  \mathbf{d}_\mathrm{b}^{n-p}  \end{array}\!\!\right] \hat{f}_b\,,\\
\begin{array}{c}~~\,\,e_b ~~\end{array}&=& ~(-1)^{p}\mathbf{tr}^{n-q}\frac{\partial \mathcal{H}}{\partial \hat{\alpha}_q}\,.
\end{eqnarray}\vspace{-0.2cm}

In contrast to (\ref{eq-38d}), in the case of the formulation (\ref{eq-38dN}), the boundary flows $\hat{f}_b$ can be considered to be freely chosen, while the boundary efforts $e_b$ are determined by the dynamics. Note that the free boundary variables are \emph{always} defined on the boundary of the dual cell complex.


\section{Physical examples}\label{Sec6}\vspace{-0.2cm}
In this section, we consider the discrete wave equation on a $2$-dimensional simplicial complex and the telegraph equations on a segment. 
\subsection{Two-dimensional wave equation}\vspace{-0.2cm}
Consider the wave equation $\mu\, \frac{\partial^2u^c}{\partial t^2}= -E \Delta u^c$, with $u^c(t,z)\in \mathbb{R}$, $z=(z_1,z_2)\in M$, where $\mu$ is the mass density, $E$ is the Young's modulus, $\Delta$ is the two-dimensional Laplace operator, and $M$ is a compact surface with a closed boundary. Throughout, the superscript $c$ designates the continuous quantities.

The energy variables are the $2$-dimensional kinetic momentum $p^c$, and the $1$-form elastic strain $\epsilon^c$. The coenergy variables are the $0$-form velocity $v^c$ and the $1$-form stress $\sigma^c$. The energy density of the vibrating membrane is $\mathcal{H}(p,\epsilon)=\frac{1}{2}\left( \epsilon^c \wedge \sigma^c + p^c \wedge v^c \right)$, where the coenergy and energy variables are related by the constitutive relations $\sigma^c=E * \epsilon^c$ and $v^c=1/\mu * p^c$. The Hodge operator here corresponds to the standard Euclidian metric on $M$. The port-Hamiltonian formulation of the vibrating membrane in full details is given in \cite{Golo}.

Let us now consider the simplicial Dirac structure underpinning the discretized two-dimensional wave equation. The energy variables of the discretized system are chosen as follows: the kinetic momentum is a dual $2$-form whose time derivative is set to be $\hat{f}_p$, the elastic strain is a primal $1$-form with time derivative corresponding to $f_q$, the coenergy variables are a primal $0$-form $e_p$ and a dual $1$-form $\hat{e}_q$. Such a formulation of the discrete wave equation is consonant with the simplicial Dirac structure (\ref{eq:Dir-prim-dual}) for the case when $p=n=2$ and $q=1$, and is given by\vspace{0.1cm}
\begin{eqnarray}\label{eq:DirWave}
 \left[\begin{array}{c}\hat{f}_p \\ {f}_q \end{array} \right] &=&\left[\begin{array}{cc} 0  & -  \mathbf{d}_\mathrm{i}^1\\  \mathbf{d}^0 & 0\end{array} \right] \left[\begin{array}{c} {e}_p \\ \hat{e}_q\end{array} \right]- \left[\begin{array}{c}  \mathbf{d}_\mathrm{b}^1 \\ 0 \end{array}\right] \hat{e}_b\,,\\
 \begin{array}{c}f_b \end{array}&=& ~\mathbf{tr}^0{e}_p\,.\notag
\end{eqnarray}\vspace{-0.2cm}

The boundary control variable is the $1$-form stress $\hat{e}_b$, while the output is the boundary velocity. The Hamiltonian of the discrete model is 
$$\mathcal{H}=\frac{1}{2}\left\langle  \epsilon \wedge E *_1\epsilon + \hat p \wedge \frac{1}{\mu} *_0^{-1} \hat p , K \right\rangle\,.$$The coenergy variables are the dual $1$-form $\hat \sigma= \frac{\partial \mathcal{H}}{\partial \epsilon}=E *_1\epsilon$ and the primal $0$-form $v =\frac{\partial \mathcal{H}}{\partial \hat{p}}=*_0^{-1}\hat p$.

The resulting port-Hamiltonian system is
\begin{eqnarray*}
 \left[\begin{array}{c} \frac{\partial \hat{p}}{\partial t} \\ \frac{\partial \epsilon}{\partial t} \end{array} \right] &=&\left[\begin{array}{cc} 0  &   \mathbf{d}_\mathrm{i}^1\\  -\mathbf{d}^0 & 0\end{array} \right]  \left[\begin{array}{cc}\frac{1}{\mu} *_0^{-1} & 0 \\0 & E *_1\end{array}\right]\left[\begin{array}{c} \hat p \\ \epsilon \end{array} \right]+ \left[\begin{array}{c}  \mathbf{d}_\mathrm{b}^1 \\ 0 \end{array}\right] \hat{e}_b\,\\
 \begin{array}{c}f_b \end{array}&=& ~\frac{1}{\mu} \mathbf{tr}^0 *_0^{-1} \hat p \,,
\end{eqnarray*}
where the operators $\mathbf{d}^0$, $\mathbf{d}_i^1$, $\mathbf{tr}^0=  ( \mathbf{d}_\mathrm{b}^1 )^\textsc{t}$, $*_1$, and $*_0^{-1}$ conform to the diagram at the end of Section~\ref{Sec3} when $n=2$.

{{
\begin{exm}
For the Dirac structure (\ref{eq:DirWave}) on the simplicial manifold $K$ given in Fig.~\ref{fig:wave}, the operators $\mathbf{d}^0$, $\mathbf{d}_i^1$, $\mathbf{tr}^0=  ( \mathbf{d}_\mathrm{b}^1 )^\textsc{t}$ are given in Example~\ref{Example1}. It is straightforward to show
\begin{eqnarray}\notag
&& \!\!\!\!\!\!\!\langle \mathbf{d}^0 e_p \wedge \hat{e}_ q ,K \rangle + \langle  e_p \wedge ( \mathbf{d}_\mathrm{i}^1 \hat{e}_ q +\mathbf{d}_\mathrm{b}^1 \hat{e}_ b ),K \rangle \\ 
&~~=& \hat{e}_b[\hat{v}_2,\hat{v}_1] f_b(v_0) + \hat{e}_b[\hat{v}_1,\hat{v}_3] f_b(v_1)\\
&~~&+\hat{e}_b[\hat{v}_3,\hat{v}_4] f_b(v_3)+\hat{e}_b[\hat{v}_4,\hat{v}_2] f_b(v_2)\,,
\notag\end{eqnarray}
what confirms that the boundary terms genuinely live on the boundary of $|K|$.
\end{exm}
}}



\subsection{Telegraph equations}
\begin{figure}
\centering
    \includegraphics[width=7cm]{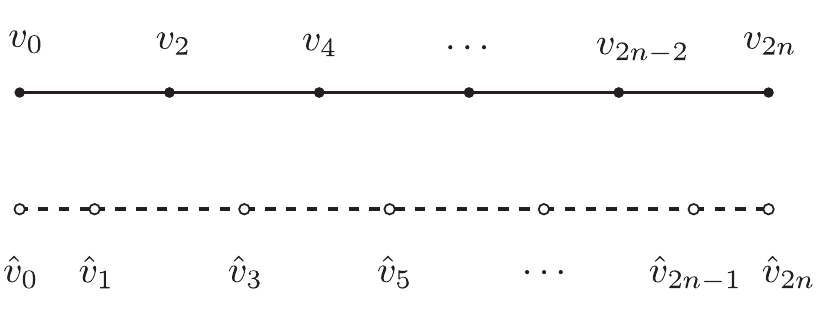}\\\vspace{-0.4cm}
  \caption{The primal $1$-dimensional simplicaial complex $K$. By construction, the nodes $\hat{v}_0$ and $\hat{v}_{2n}$ are added to the boundary to insure that $\partial (\star K)=\star (\partial K)$. }\label{fig:RDcomp1}
\end{figure}

We consider an ideal lossless transmission line on a $1$-dimensional simplicial complex given in Fig.~\ref{fig:RDcomp1}. The energy variables are the charge density ${q}\in \Omega_d^1(K)$, and the flux density $\hat{\phi}\in\Omega_d^1(\star  K)$, hence $p=q=1$. The Hamiltonian representing the total energy stored in the transmission line with distributed capacitance $C$ and distributed inductance $\hat{L}$ is\vspace{-0.1cm}
\begin{eqnarray}
\mathcal{H}= \left \langle \frac{1}{2C} {q} \wedge *_1 {q}  +  \frac{1}{2\hat{L}}  \hat{\phi} \wedge *_0^{-1} \hat{\phi}  ,K\right\rangle \,,
\end{eqnarray}
where $*_0$ and $*_1$ are the discrete diagonal Hodge operators that relate the appropriate cochains according to the following schematic diagram\vspace{-0.3cm}
\begin{equation*}
\left.\begin{array}{ccccc} \!\!\! \Omega_d^0(\partial K)\!\!\! & \!\!\! \xleftarrow{~\mathbf{tr}^0~} \!\!\! & \!\!\! \Omega_d^0(K)\!\!\! & \!\xrightarrow{~\mathbf{d}^0~}\!\! &  \Omega_d^1(K)\!\!\! \\ 
\!\!\!\downarrow\! {*_\mathrm{b}} \!\!\!& \!\!\! \!\!\! &\!\!\!  \downarrow\!{*_0} \!\!\!&\!\!\!\!\!\!  & \!\!\! \downarrow\!{*_1} \!\!\! \\ 
\! \Omega_d^0(\partial(\star  K))\! & \xrightarrow{~\mathbf{d}_\mathrm{b}^0~} &  \Omega_d^1(\star_\mathrm{i} K)  & \!\xleftarrow{{~\mathbf{d}}_\mathrm{i}^0~}\! & ~ \Omega_d^0(\star_\mathrm{i} K)\,,\!\!\! \end{array}\right.\,\vspace{-0.4cm}
\end{equation*}
where $\star_\mathrm{b}$ is the identity.

The co-energy variables are: $\hat{e}_p=\hat{\frac{\partial \mathcal{H}}{\partial{q}}}=*\frac{{q}}{C}=\hat{V}$ representing voltages and ${e}_q=\frac{\partial \mathcal{H}}{\partial \hat{\phi }}=*\frac{\hat{\phi}}{\hat{L}}=I$ currents. Selecting ${f}_p=-\frac{\partial{q}}{\partial  t}$ and $\hat{f}_q=-\frac{\partial \hat{\phi}}{\partial  t}$ leads to the port-Hamiltonian formulation of the telegraph equations
\begin{eqnarray}\label{eq:Dir-prim-dualMET}
\left[\begin{array}{c} - \frac{\partial {q} }{\partial t} \\ -\frac{\partial \hat{\phi} }{\partial t}\end{array}\right]&=&\left[\begin{array}{cc} 0 & \mathbf{d}^0 \\  \mathbf{d}_\mathrm{i}^0 & 0\end{array}\right] \left[\begin{array}{c} *_1\frac{q}{C} \\  *_0^{-1}\frac{ \hat{\phi} }{\hat{L}}  \end{array}\right]+\left[\begin{array}{c} 0 \\  \mathbf{d}_\mathrm{b}^0  \end{array}\right] \hat{f}_b\,\\
\begin{array}{c}~~\,\,e_b ~~\end{array} &=& ~ - \mathbf{tr}^0*_0^{-1} \frac{\hat{\phi}}{\hat{L}} \,,\notag
\end{eqnarray}where $\hat{f}_b$ are the input voltages and $e_b$ are the output currents.

In the case we want to have the electrical currents as the inputs, the charge and the flux densities would be defined on the dual mesh and the primal mesh, respectively. Instead of the port-Hamiltonian system in the form (\ref{eq:Dir-prim-dualMET}), the discretized telegraph equations would be in the form (\ref{eq-38d}). The charge density is defined on the dual cell complex as $\hat{q}\in \Omega_d^1(\star_\mathrm{i} K)$ and the discrete flux density is $\phi \in \Omega_d^1(K)$. The finite-dimensional port-Hamiltonian system is of the form
\begin{eqnarray}\label{eq:Dir-prim-dualMETcurrent}
\left[\begin{array}{c}- \frac{\partial \hat{q} }{\partial t} \\ -\frac{\partial \phi }{\partial t}\end{array}\right]&=&\left[\begin{array}{cc} 0 & \mathbf{d}_\mathrm{i}^0 \\  \mathbf{d}^0 & 0\end{array}\right] \left[\begin{array}{c} *_0^{-1} \frac{\hat{q}}{\hat{C}} \\  *_1\frac{ \phi }{L}  \end{array}\right] +\left[\begin{array}{c}   \mathbf{d}_\mathrm{b}^0 \\ 0 \end{array}\right] \hat{e}_b\,\\\notag
\begin{array}{c}~~\,\,f_b ~~\end{array} & = & ~ - \mathbf{tr}^0 *_0^{-1} \frac{\hat{q}}{\hat{C}}  \,,
\end{eqnarray}
where $\hat{e}_b$ are the input currents and $f_b$ are the output voltages.

\vspace{0.15cm}
The exterior derivative $\mathbf{d}^0: \Omega_d^0(K)\rightarrow \Omega_d^1(K)$ is the transpose of the incidence matrix of the primal mesh. The discrete derivative $\mathbf{d}_\mathrm{i}^0: \Omega_d^0(\star_\mathrm{i} K)\rightarrow \Omega_d^1(\star_\mathrm{i} K)$ in the matrix notation is the incidence matrix of the primal mesh. Thus, we have
\begin{eqnarray}\label{eq-6Nat1}
-(\mathbf{d}_\mathrm{i}^0)^\textsc{t}&=&\mathbf{d}^0=\left[\begin{array}{rrrrrr}-1 & 1 & 0 & \cdots & 0 & 0 \\0 & -1 & 1 & \cdots & 0 & 0 \\ \, & \, & \, & \ddots & \, & \, \\0 & 0 & 0 & \cdots & -1 & 1\end{array}\right]\,,\\
\mathbf{tr}^0&=&(\mathbf{d}_\mathrm{b}^0)^\textsc{t}=\left[\begin{array}{rrrrrr}-1 & 0 & 0 & \cdots & 0 & 0 \\0 & 0 & 0 & \cdots & 0 & 1\end{array}\right]\,.
\end{eqnarray}


\vspace{0.1cm}
\begin{rmk}The discrete analogue of the Stokes-Dirac structure obtained in \cite{Golo} is a finite-dimensional Dirac structure, but not a Poisson structure. The implication of this on the physical realization is that, in contrast to our results, the transmission line in the finite-dimensional case is not only composed of inductors and capacitors but also of transformers.\end{rmk}

\begin{figure}
\centering
    \includegraphics[width=7.8cm]{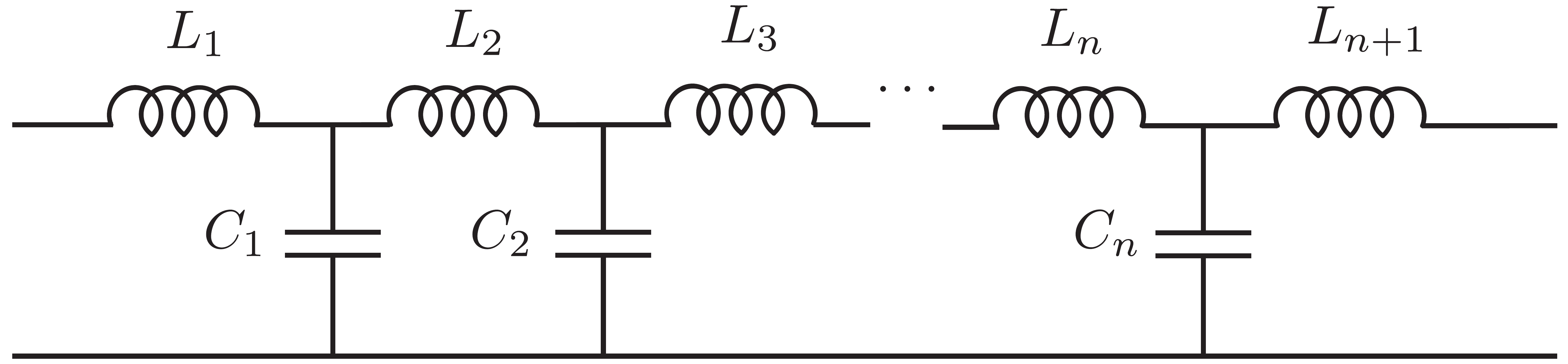}\\\vspace{-0.0cm}
  \caption{The finite-dimensional approximation of the lossless transmission line when the inputs are voltages and the outputs are currents. The inductances $L_1,\ldots, L_{n+1}$ are the values that the discrete distributed inductance $\hat{L}$ takes on the simplices $[\hat{v}_0,\hat{v}_1],\ldots, [\hat{v}_{2n-1},\hat{v}_{2n}]$; the capacitances $C_1,\ldots C_n$ are the values $C$ takes on $[v_0,v_2],\ldots, [v_{2n-2},v_{2n}]$.}\label{fig:LC1}
  \end{figure}

\begin{figure}
\centering
    \includegraphics[width=7.8cm]{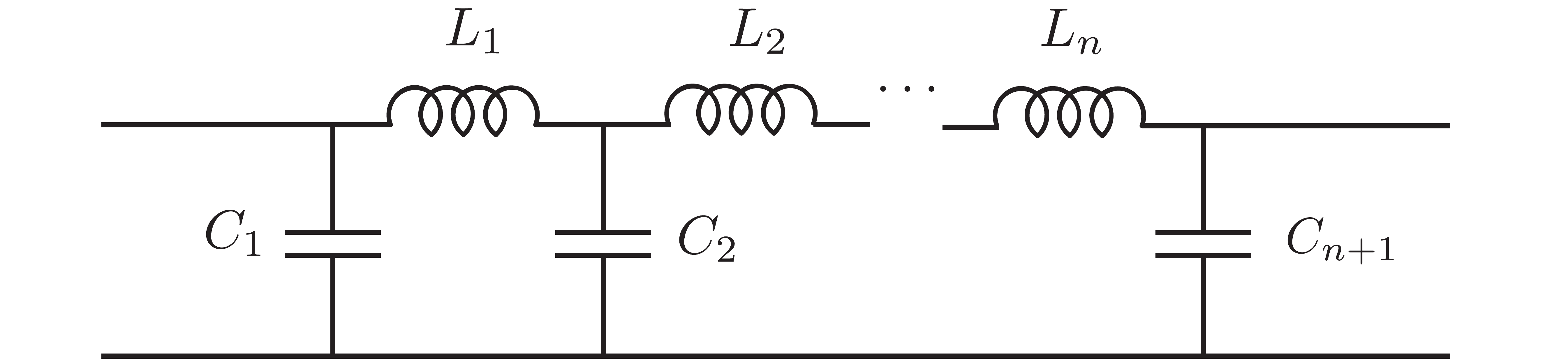}\\\vspace{-0.0cm}
  \caption{The finite-dimensional approximation of the lossless transmission line when the inputs are currents and the outputs are voltages. The inductances are: $L_1=\int_{[v_0,v_2]}L^c=L([v_0,v_2])$, $L_2=\int_{[v_2,v_4]}L^c=L([v_2,v_4])$, $\ldots$, $L_n=\int_{[v_{2n-2},v_{2n}]} L^c =L([v_{2n-2},v_{2n}])$; the values of capacitors are: $C_1=\int_{[\hat{v}_0,\hat{v}_1]}C^c=\hat{C}([\hat{v}_0,\hat{v}_1])$, $C_2=\int_{[\hat{v}_1,\hat{v}_3]}C^c=\hat{C}([\hat{v}_1,\hat{v}_3])$, $C_3=\int_{[\hat{v}_3,\hat{v}_5]}C^c=\hat{C}([\hat{v}_3,\hat{v}_5])$, $\ldots$, $C_{n+1}=\int_{[\hat{v}_{2n-1},\hat{v}_{2n}]}C^c=\hat{C}([\hat{v}_{2n-1},\hat{v}_{2n}])$.}\label{fig:LC2}
  \end{figure}

The physical realizations of the port-Hamiltonian systems (\ref{eq:Dir-prim-dualMET}) and (\ref{eq:Dir-prim-dualMETcurrent}) are given on  Fig.~\ref{fig:LC1} and Fig.~\ref{fig:LC2}, respectively. Stabilization of either of those systems is easily achieved by terminating boundary ports with resistive elements, what is a practical application of the passivization explained in Remark~\ref{rmk:pass}.

\begin{rmk}The accuracy of the proposed method is $1/n$ (see \cite{SeslijaArXiv}).
\end{rmk}

\section{Conservation laws}\label{Sec7}
Let us consider the existence of conservation laws and structural invariants for the port-Hamiltonian systems on simplicial complexes. 


\subsection{Finite-dimensional invariants}
The following proposition gives the conditions for the existence of conservation laws in the discrete setting.

\begin{prop}Consider the port-Hamiltonian system (\ref{eq-38d}). Let $(\hat\alpha_p,\alpha_q)\mapsto C(\hat\alpha_p,\alpha_q)$ be a real-valued function. Then
\begin{eqnarray}\label{eq:Prop_1}
\frac{\partial C}{\partial \hat\alpha_p}&\in \mathrm{ker}\, \mathbf{d}^{n-p}\\\label{eq:Prop_2}
\hat{\frac{\partial C}{\partial \alpha_q}}&\in \mathrm{ker}\, \mathbf{d}_\mathrm{i}^{n-q}\,,
\end{eqnarray}
iff $C$ is a conservation law for the port-Hamiltonian system (\ref{eq-38d}) satisfying
\begin{equation}\label{eq:Prop_3}
\frac{\mathrm{d} C}{\mathrm{d}t}= \left(f_b^C\right)^\textsc{t} \hat e_b\,,
\end{equation}
where $f_b^C=-(-1)^{q(p+1)} \mathbf{tr}^{n-p}\frac{\partial C}{\partial \hat\alpha_p}$.
\end{prop}\vspace{-0.2cm}
\begin{pf}
Differentiating $C$ along the flow of the system (\ref{eq-38d}), we have
\begin{eqnarray}\label{eq:Prop_4}\notag
\frac{\mathrm{d} C}{\mathrm{d}t}&=&\left\langle \frac{\partial {C}}{\partial \hat{\alpha}_p} \wedge \frac{\partial \hat{\alpha}_p}{\partial t} + \hat{\frac{\partial C}{\partial \alpha_q}}\wedge \frac{\partial \alpha_q }{\partial t},K\right\rangle\\\notag
&=& (-1)^{pq}\frac{\partial^\textsc{t} {C}}{\partial \hat{\alpha}_p} \left( \mathbf{d}_\mathrm{i}^{n-q}\hat{\frac{\partial \mathcal{H}}{\partial \alpha_q}} +\mathbf{d}_\mathrm{b}^{n-q}\hat e_b\right)\\
&~&-(-1)^{q(n-q)} \left(\mathbf{d}^{n-q} \frac{\partial \mathcal{H}}{\alpha_q }\right)^\textsc{t} \hat{\frac{\partial {C}}{\partial \alpha_q}} \\\notag
&=&(-1)^{q(p+1)} \left( \mathbf{d}^{n-p} \frac{\partial {C}}{\partial \hat{\alpha}_p} \right)^\textsc{t} \hat{\frac{\partial \mathcal{H}}{\partial \alpha_q}}\\\notag
&~&-(-1)^{q(p+1)} \left( \mathbf{tr}^{n-p} \frac{\partial {C}}{\partial \hat{\alpha}_p} \right)^\textsc{t} \hat e_b\\\notag
&~&+ (-1)^{pq} \left( \mathbf{d}_\mathrm{i}^{n-q} \hat{\frac{\partial {C}}{\partial \alpha_q}} \right)^\textsc{t} {\frac{\partial {\mathcal{H}}}{\partial \hat \alpha_p}}\,,
\end{eqnarray}
where we have used the fact that $\mathbf{d}_\mathrm{i}^{n-q}=(-1)^q (\mathbf{d}^{n-p})^\textsc{t}$ and $\mathbf{d}_\mathrm{b}^{n-q}=(-1)^{n-p}(\mathbf{tr}^{n-p})^\textsc{t}$. Furthermore, regardless of $\mathcal{H}$, the result (\ref{eq:Prop_3}) follows iff (\ref{eq:Prop_1}) and (\ref{eq:Prop_2}) hold. \qed
\end{pf}

\begin{rmk}
If either $\hat e_b=0$ or $f_b^C=0$, the quantity $C$ satisfying (\ref{eq:Prop_1}) and (\ref{eq:Prop_2}) is a conserved quantity---a Casimir function.
\end{rmk}

\subsection{One-dimensional domain}
An interesting case for which it is possible explicitly to solve (\ref{eq:Prop_1}) is when $p=n$. The matrix $\mathbf{d}^0$ is nothing but the transpose of the incidence matrix $\partial_1$, from the set of edges to the set of vertices, on a connected graph. It is a well-known property of any incidence matrix $\partial_1$ that $\mathrm{ker}\, \partial_1^\textsc{t}=\mathrm{span}\, \mathbf{1}$, where $\mathbf{1}$ stands for the vector with all elements equal $1$. A direct consequence of this is that $\frac{\partial C}{\partial \hat \alpha_p}= \mathbf{1}$ up to a multiplicative constant.\vspace{-0.2cm}

In the one-dimensional case the null space of $\mathbf{d}_\mathrm{i}^0$ is trivial, cf. (\ref{eq-6Nat1}), what allows us to explicitly express the conservation law.\vspace{-0.2cm}

\begin{cor}
Consider the port-Hamiltonian system (\ref{eq-38d}), with $p=q=n=1$, on a one-dimensional simplicial manifold given on Figure~\ref{fig:RDcomp1}. The quantity $C_p=\mathbf{1}^\textsc{t}\hat\alpha_p=\hat\alpha_p([\hat v_0,\hat v_1]) + \sum_{k=1}^{n-1} \hat\alpha_p([\hat v_{2k-1},\hat v_{2k+1}]) +\hat\alpha_p([\hat v_{2n-1},\hat v_{2n}])$ satisfies the balance law
\begin{equation}
\frac{\mathrm{d} C_p}{\mathrm{d} t}= \hat e_b(\hat v_0)- \hat e_b(\hat v_{2n}).
\end{equation}
\end{cor}\vspace{-0.3cm}

In case of the telegraph equations on the segment $M=[0,1]$, the total charge $C_q^c=\int_0^1 q^c(t,z)\mathrm{d}z$ as well as the total magnetic flux $C_\phi^c=\int_0^1 \phi^c(t,z)\mathrm{d} z$ are both conservation laws. In the discrete setting, the \emph{only} conservation law for the system (\ref{eq:Dir-prim-dualMETcurrent}) is the total charge $C_q=\mathbf{1}^\textsc{t} \hat q$ whose derivative along the admissible trajectories is $\frac{\mathrm{d} C_q}{\mathrm{d} t}=\hat e_b(\hat v_0)-\hat e_b(\hat v_{2n})$. Similarly, the total flux $C_\phi=\mathbf{1}^\textsc{t} \hat \phi$ in the system (\ref{eq:Dir-prim-dualMET}) satisfies the balance law $\frac{\mathrm{d} C_\phi}{\mathrm{d} t}=\hat f_b(\hat v_0)- \hat f_b(\hat v_{2n})$, where $\hat f_b(\hat v_0)$ and $\hat f_b(\hat v_{2n})$ are input currents. These result differ from those presented in \cite{Macchelli}, where both the total flux and total charge are conserved.

\vspace{-0.1cm}
\section{Energy-Casimir method}\label{Sec:ECmethod}\vspace{-0.2cm}
Consider the interconnection of (\ref{eq-38d}) with the (possibly nonlinear) integrator\vspace{-0.2cm}
\begin{eqnarray}
\frac{\mathrm{d}\zeta}{\mathrm{d} t}&=& \mathbf{g}_c\, u_c\\
y_c&=& \mathbf{g}_c^\textsc{t}\,\frac{\partial H_c}{\partial \zeta}\,,
\end{eqnarray}\vspace{-0.0cm}
where $\zeta\in {\mathbb{R}}^m$, $\mathbf{g}_c\in \mathbb{R}^{m\times N_b}$ with $N_b=\mathrm{dim} \,\Omega_d^{n-q}(\partial (\star K))$, input $u_c$, output $y_c$, and $\zeta \mapsto H_c(\zeta)$ the controller's Hamiltonian. The interconnection is power-preserving with $u_c=f_b$ and $e_b=-y_c$. The composition is the port-Hamiltonian system in the form
\begin{equation}\label{eq:EnergyCasimirClosed}
\!\!\!\left[\!\begin{array}{c}\frac{\partial\hat \alpha_p}{\partial t} \\\frac{\partial \alpha_q}{\partial t} \\\frac{\mathrm{d}\zeta}{\mathrm{d} t}\end{array}\!\right] \!\!=\!\! \left[\!\begin{array}{ccc} 0  &\!\!\!\! \!\!\!(-1)^{r-1}  \mathbf{d}_\mathrm{i}^{n-q} &(-1)^{r} \mathbf{d}_\mathrm{b}^{n-q} \mathbf{g}_c^\textsc{t}  \\  \mathbf{d}^{n-p} & 0 & 0\\
(-1)^{p}\mathbf{g}_c\,\mathbf{tr}^{n-p} & 0 &
\end{array}\! \right] \left[\!\begin{array}{c} {\frac{\partial {H}_{\mathrm{cl}}}{\partial \hat{\alpha}_p}} \\\hat{\frac{\partial {H}_{\mathrm{cl}}}{\partial \alpha_q}}\\ \frac{\partial H_{\mathrm{cl}}}{\partial \zeta}\end{array}\!\right],
\end{equation}
with $(\hat\alpha_p,\alpha_q,\zeta)\mapsto H_{\mathrm{cl}}(\hat \alpha_p,\alpha_q,\zeta)$ is the closed-loop Hamiltonian $H_\mathrm{cl}(\hat \alpha_p,\alpha_q,\zeta)=\mathcal{H}(\hat \alpha_p,\alpha_q)+H_c (\zeta)$.

The energy shaping for the system (\ref{eq:EnergyCasimirClosed}) is achieved by restricting the behavior of (\ref{eq:EnergyCasimirClosed}) to a certain subspace \cite{L2gain}. To this end, we look at the Casimir functions of the closed-loop system.

\begin{prop}
The real-valued function $(\hat\alpha_p,\alpha_q,\zeta)\mapsto C(\hat\alpha_p,\alpha_q,\zeta)$ is a Casimir function of the closed system (\ref{eq:EnergyCasimirClosed}) iff
\begin{eqnarray}\label{eq:EnergyCasimirClosedCas}\notag
\frac{\partial C}{\partial \hat\alpha_p}&\in& \mathrm{ker}\, \mathbf{d}^{n-p}\cap \mathrm{ker} \left(\mathbf{g}_c\,\mathbf{tr}^{n-p}\right)\\
\left[\begin{array}{c}\hat{\frac{\partial C}{\partial \alpha_q}} \\\frac{\partial C}{\partial \zeta}\end{array}\right]&\in& \mathrm{ker} \left[\begin{array}{cc}\mathbf{d}_\mathrm{i}^{n-q} & (-1)^{n-q} \mathbf{d}_\mathrm{b}^{n-q} \mathbf{g}_c^\textsc{t}\end{array}\right]\,.
\end{eqnarray}
\end{prop}
\begin{pf}
Solving $\frac{\mathrm{d}}{\mathrm{d} t}C(\hat \alpha_p,\alpha_q,\zeta)=0$ irrespective of $H_\mathrm{cl}$ directly leads to (\ref{eq:EnergyCasimirClosedCas}). \qed
\end{pf}

\begin{rmk}Since the structural matrix of the port-Hamiltonian system (\ref{eq-38d}) is \emph{not} of full rank in case when $\mathbf{g}_c$ is identity, not all Casimirs of (\ref{eq:EnergyCasimirClosed}) are of the form $C(\hat \alpha_q,\alpha_q,\zeta)=S_i(\hat \alpha_p,\alpha_q)-\zeta_i$, $i=1,\ldots,m$.
\end{rmk}

\begin{rmk}In case when $p=q=m=n=1$ and $\mathbf{g}_c=[1,1]$, the \emph{only} Casimir for the system (\ref{eq:EnergyCasimirClosed}) is $\mathbf{1}^\textsc{t} \alpha_q+ \zeta$.
\end{rmk}


\section{Final remarks}{{
The explicit simplicial discretization treated in this paper leads to the standard input-output port-Hamiltonian systems without algebraic constraints. The analysis and the control synthesis for such systems belong to the realm of standard finite-dimensional systems.

In the last section, we looked at a simple control strategy for the energy shaping of discretized port-Hamiltonian systems. This attempt has only scratched the surface of a very important problem. Since the discretized model assumes a port-Hamiltonian structure, much more elaborate schemes for the control of port-Hamiltonian systems can be applied. A nontrivial problem in this regard would be to design a controller for the discretized model and then test it on the continuous model and obtain the bounds of the discrepancy norm between the two behaviors. Some initial work has already been done in this vein, however, mostly pertaining to the systems on a one-dimensional spatial domain \big(see \cite{Macchelli,Voss2} and references quoted therein\big). In higher dimensions, the interconnection of the finite controller and the infinite-dimensional plant would be naturally realized through the interface of the simplicial triangulation of the boundary. Gauging the input-output errors and energy shaping of the closed-loop systems we plan to explore in forthcoming publications.}}





\vspace{-0.0cm}

\end{document}